\documentclass[10pt]{article}

\usepackage[hidelinks, colorlinks=true]{hyperref}

\usepackage[letterpaper, top=1in, bottom=1in, right=0.8in, left=0.9in]{geometry}

\usepackage{color}
\usepackage{graphicx}
\usepackage{url}
\usepackage{subfig}
\usepackage{caption}
\usepackage{float}
\usepackage{multicol, lipsum}
\usepackage{amsmath}
\usepackage{paralist} % for compactitem
\usepackage{titling}

\begin{document}

\pagenumbering{arabic}  % {roman}

\title{\noindent\rule{\textwidth}{4pt}\\\vspace{12pt}\textbf{Generator From Edges: Reconstruction of Facial Images}\\
\noindent\rule{\textwidth}{1pt}}

\date{\vspace{-8ex}} %remove date
%\date{}

\maketitle

%\author{Nao Takano, Gita Alaghband}
\begin{center}
\begin{tabular}{ c c c }
 \textbf{Nao Takano} &  \hspace{1.5cm} & \textbf{Gita Alaghband} \\ 
 Computer Science and Engineering & & Computer Science and Engineering \\  
 University of Colorado Denver & & University of Colorado Denver     
\end{tabular}
\end{center}

\vspace{24 pt}

\begin{abstract}
\noindent
Applications that involve supervised training require paired images.  Researchers of single image super-resolution (SISR) create such images by artificially generating blurry input images from the corresponding ground truth.   Similarly we can create paired images with the canny edge.  We propose Generator From Edges (GFE) [Figure \ref{figImages}].  Our aim is to determine the best architecture for GFE, along with reviews of perceptual loss \cite{Johnson2016, Ledig2016}.  To this end, we conducted three experiments.  First, we explored the effects of the adversarial loss often used in SISR.  In particular, we uncovered that it is not an essential component to form a perceptual loss.  Eliminating adversarial loss will lead to a more effective architecture from the perspective of hardware resource.  It also means that considerations for the problems pertaining to generative adversarial network (GAN) \cite{Goodfellow2014}, such as mode collapse, are not necessary.  Second, we reexamined VGG loss and found that the mid-layers yield the best results.  By extracting the full potential of VGG loss, the overall performance of perceptual loss improves significantly.  Third, based on the findings of the first two experiments, we reevaluated the dense network to construct GFE.  Using GFE as an intermediate process, reconstructing a facial image from a pencil sketch can become an easy task.
\end{abstract}

\vspace{32 pt}

\begin{multicols}{2}

%%%%%%%%%%%%%%%%%%%%%%%%%%%%%%%%%%%%%%%%%%%%%%%%%%%%%%%%%%%%%%%%%%%%%%%%%%%%%%%%
\section{Introduction}
%%%%%%%%%%%%%%%%%%%%%%%%%%%%%%%%%%%%%%%%%%%%%%%%%%%%%%%%%%%%%%%%%%%%%%%%%%%%%%%%

While there have been quite a few methods and proposals for single image super-resolution (SISR), few applications exist for reconstructing an original face from the corresponding edge image.  The techniques used in our Generator From Edges (GFE) are variations of those used in SISR.  In SISR, there are roughly two categories.  The first is by way of the perceptual loss that includes adversarial loss, which requires a generative adversarial network.  The second omits adversarial loss.  GFE belongs to the second category.

GAN, used in the first category, consists of two networks; a generator and a discriminator, and adversarial loss, used in a supervised setting, is derived from the discriminator.

\begin{figure}[H]
\centering
\fbox{\includegraphics[width=.45\linewidth]{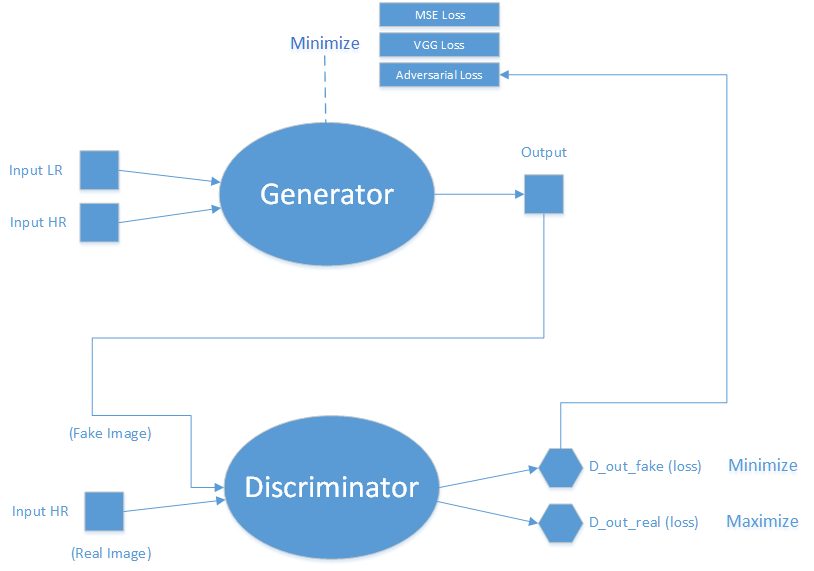}} % using fbox because we need a border
\hspace{.005cm}   %\hfill
\fbox{\includegraphics[width=.45\linewidth]{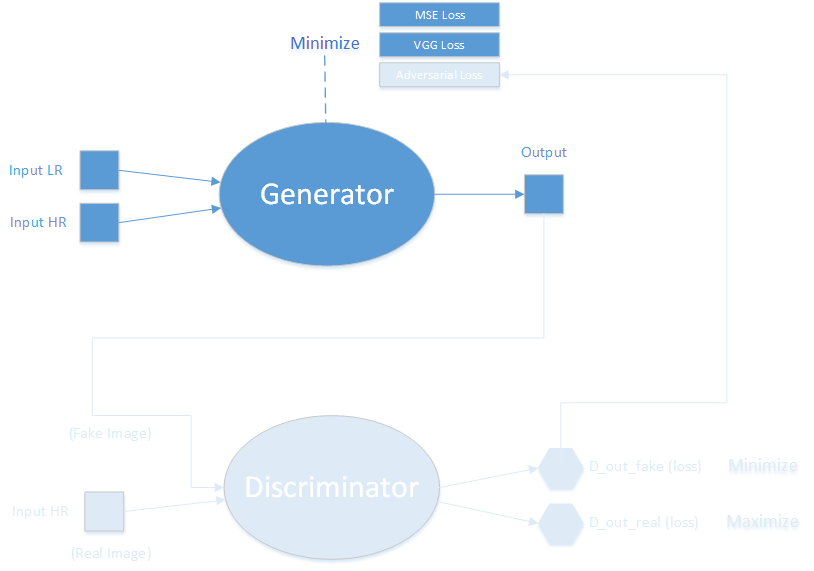}}
\caption{Left: SRGAN diagram, Right: GFE, the proposed architecture -- many of the elements go away}
\label{figSRGAN}
\end{figure}

We focus on three perceptual losses that have larger impact on the overall performance; adversarial loss, MSE loss, and VGG loss.  Removing adversarial loss results in a simpler architecture, enabling us to eliminate the discriminator.  Figure \ref{figSRGAN} depicts the differences between SRGAN \cite{Ledig2016}, one of the most influential works for SISR, and our proposed GFE.

In general, the larger the neural network model, the better the outcome we expect. This is true in the accuracy of classification as well as in the synthesis of images.  But as the size of the input image and the complexity of the synthesis grow, we come across problems with ``instabilities specific to large-scale GANs" \cite{Brock2018LargeSG} and the rate of contribution by the discriminator diminishes.  In addition, the discriminator requires a large memory footprint.  In order to use a large network for training with the limited amount of memory for hardware such as GPU, the question arises as to how much the discriminator contributes to the outcome of the synthesis.  We observed very little, if any, positive effect using the adversarial loss for GFE.  If we do away with the discriminator, we can free up its otherwise occupied GPU memory, making it possible to construct a larger generator.  Moreover, using only two loss functions (image loss and VGG loss -- Section \ref{secPeceptualLosses}) permits easier settings for hyper-parameters.  We attempt to measure the effect of the discriminator and perform image synthesis without it.

This is all achieved without sacrificing the fidelity of the outputs.  In the sections that follow, we present our contributions by describing details for the three experiments conducted in this study.  We define Generator From Edges to be a generator (in the same sense as the one used in GAN, but without discriminator) for an application that restores images from their corresponding canny edge.  All the experiments train with pairs of images; a ground truth and the corresponding single-channel (grayscale) edge image.  The edge images are created by running the OpenCV Canny function from CelebA (Experiment 1) and CelebA-HQ \cite{KarrasHQ} datasets (Experiments 2 and 3). 

\end{multicols}

%\newpage
\begin{multicols}{1}
\begin{figure}[H]
\centering
\subfloat{\includegraphics[width=.5\textwidth]{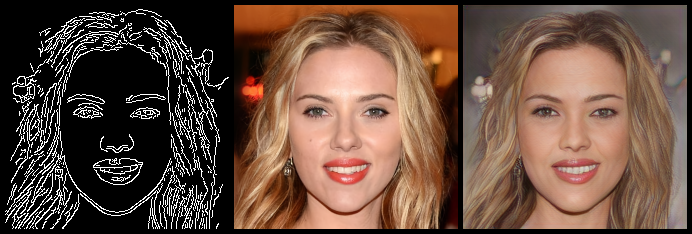}}
\hspace*{0.1cm}   %\hfill
\subfloat{\includegraphics[width=.5\textwidth]{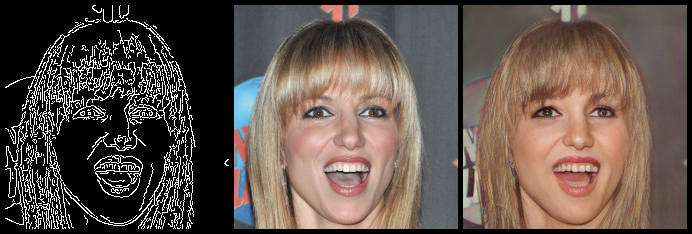}}\\
\vspace{1pt}
\subfloat{\includegraphics[width=.5\textwidth]{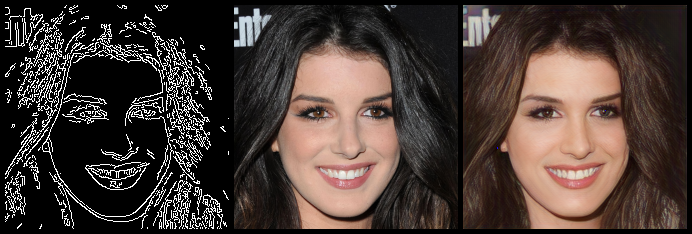}}
\hspace*{0.1cm}   %\hfill
\subfloat{\includegraphics[width=.5\textwidth]{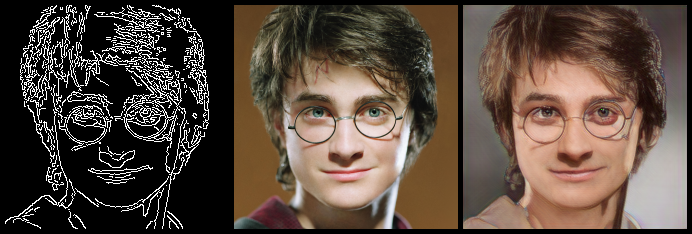}}\\
%\vspace{6pt}
\subfloat{\includegraphics[width=.5\textwidth]{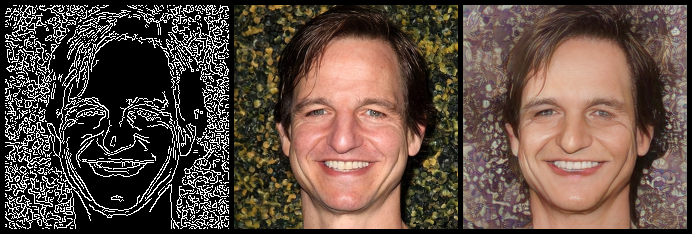}}
\hspace*{.1cm}   %\hfill
\subfloat{\includegraphics[width=.5\textwidth]{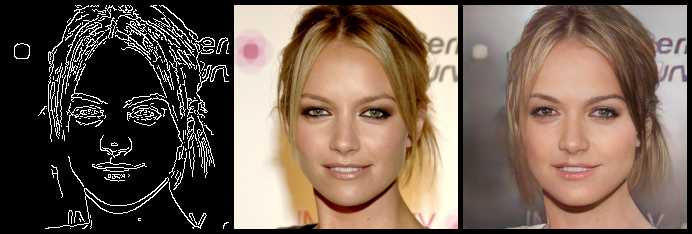}}\\
%\vspace{6pt}
\subfloat{\includegraphics[width=.5\textwidth]{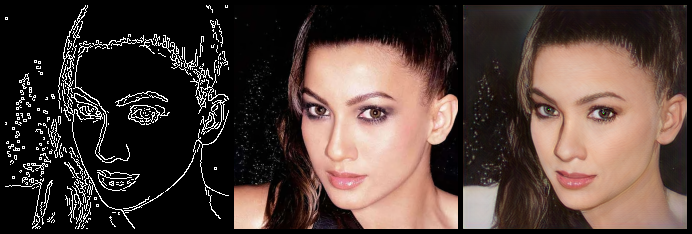}}
\hspace*{.1cm}   %\hfill
\subfloat{\includegraphics[width=.5\textwidth]{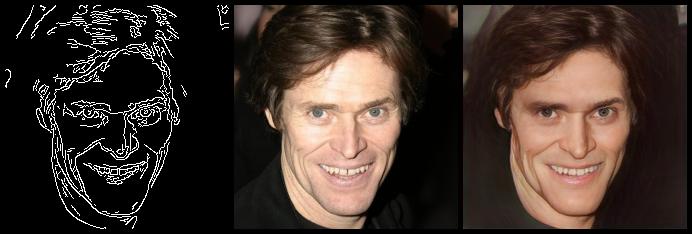}}

\caption{From Left: Input, GT, Output; CelebA-HQ \cite{KarrasHQ} contains 30,000 images and after subtracting 2 outliers,\\ we used 28,998 for training and 1,000 for testing.  These are sample outcomes from the testing set.}
\label{figImages}
\end{figure}
\end{multicols}

%\hspace*{0.4cm}
%\subfloat {\includegraphics[width=.45\textwidth]{Sketches/fourImages_Sketches_f1-001-01-sz1_150-300.png}}
%\hspace*{0.3cm}

\newpage
\begin{multicols}{2}

\begin{itemize}

\item Experiment 1 ---  Effect of Adversarial Loss

Using CelebA dataset, we measured the effectiveness of adversarial loss for both SISR and the synthesis from canny images.  The larger the images and more complex the task becomes, the less adversarial loss contributes to the outcome.  (Section \ref{secAdvLoss})

\item Experiment 2 --- VGG loss

VGG loss is another element of perceptual loss.  The absence of adversarial loss leaves our perceptual loss more reliant on the VGG loss \cite{Ledig2016, Wang2018}.  We used CelebA-HQ dataset with the image size $224 \times 224$.  Using middle layers is more effective than using the last layers. (Section \ref{secVGG})

\item Experiment 3 --- Dense Network

In pursuit of the best quality, we investigate and propose the network architecture for GFE.  The specific focus is on the effectiveness of dense connections in the network.  Each residual block should have exactly one batch normalization layer, and skip connections are ineffective. (Section \ref{secGFE})

\end{itemize}

%%%%%%%%%%%%%%%%%%%%%%%%%%%%%%%%%%%%%%%%%%%%%%%%%%%%%%%%%%%%%%%%%%%%%%%%%%%%%%%%
\section{Related Works}
%%%%%%%%%%%%%%%%%%%%%%%%%%%%%%%%%%%%%%%%%%%%%%%%%%%%%%%%%%%%%%%%%%%%%%%%%%%%%%%%
\label{secRelatedWorks}

Reconstruction from edge images is a derivative of the SISR problem, thus it is imperative to consider SISR first.  Since the inception of GAN \cite{Goodfellow2014}, the technique to use a discriminator was adopted by SRGAN \cite{Ledig2016}, whose influences and follow-up research \cite{Wang2018, Vu2018PerceptionEnhancedIS, Liu2018AnAA, Feng2019SuppressingMO} are the inspiration for our work.
For human facial synthesis, whereas we generate a photo from edges (sketch), other research generates a sketch from an input photo \cite{Yu2017TowardsRF, Zhang2018RobustFS, Zhu2017DeepGF}.  Huang et al. \cite{Huang2017BeyondFR} made a frontal view synthesis from profile images.  Di et al. \cite{Di2017GPGANGP} used a combination of U-Net and DenseNet to generate faces from landmark keypoints.  Li et al. \cite{Li2017GenerativeFC} as well as Jo and Park \cite{Jo2019SCFEGANFE} generated facial images with a partial reconstruction from sketches.  In terms of application, our work is related to Lu et al. \cite{Lu2017ImageGF} and Yu et al. \cite{Yu2019ImprovingFS}.  The former uses Contextual GAN, where input photos and images are trained in semi-supervised fashion.  The latter uses a Conditional CycleGAN to address the heterogeneous nature of photos and sketches.  StarGAN \cite{Choi2017StarGANUG} offers an impressive multi-domain image-to-image translation on human faces.  Liu et al. \cite{Liu2019ComponentSP} uses SR for human faces.  In the general domains to generate images from sketches other than human faces, substantial advancements have been made \cite{Pathak2016ContextEF, Chen2018SketchyGANTD, Wu2017GPGANTR, Zhao2018GuidedII, Yu2018FreeFormII, Wang2019MultiInstanceST}.  More recently, Mask based image synthesis on human faces \cite{Gu2019MaskGuidedPE, Lee2019MaskGANTD} is gaining attention.  Also, Qian et al. \cite{Qian2019MakeAF} propose Additive Focal Variational Auto-encoder (AF-VAE) for facial manipulation.

Going back a few years, a wide applicability of image-to-image translation in the supervised setting was proposed by Isola et al. \cite{Isola2016} with GAN.  Our work follows this line of research; but instead of adversarial loss, we use VGG loss.

Chen et al. \cite{ChenKoltun} demonstrated a similar applicability to \cite{Isola2016} but with a Cascaded Refinement Network, which starts with a low resolution module and doubles its size for consecutive modules.  For SISR, VDSR \cite{Kim2015AccurateIS} and SRResNet \cite{Ledig2016} were notable architectures prior to SRGAN.  Lim et al. \cite{Lim2017} used a multi-scale model (EDSR) that enables flexible input image size, which also reduces the number of parameters.  Tong et al. \cite{Tong2017} applied dense skip connections in a very deep network to boost the SR reconstruction.  Mei et al. \cite{Mei2019DeepRR} used a multi-frame network to generate more than one output and fused them into a single output.  Ma, et al. \cite{Ma2019AMA} replaced simple skip connections with the connection nodes and proposed a multi-level aggregated network (MLAN).  The research presented in these papers successfully synthesized images without using a GAN, which led us to ask ourselves: If we can create images without GAN, then how much does a GAN contribute to the outcome?  If we drop it from our system altogether, what would be gained?  We examine these topics in the context of the synthesis/reconstruction of an image of a human face.

%%%%%%%%%%%%%%%%%%%%%%%%%%%%%%%%%%%%%%%%%%%%%%%%%%%%%%%%%%%%%%%%%%%%%%%%%%%%%%%%
\section{Perceptual Loss Functions}
%%%%%%%%%%%%%%%%%%%%%%%%%%%%%%%%%%%%%%%%%%%%%%%%%%%%%%%%%%%%%%%%%%%%%%%%%%%%%%%%
\label{secPeceptualLosses}

Perceptual loss functions were first defined by Johnson, et al. \cite{Johnson2016} and adopted by Ledig, et al. \cite{Ledig2016}.  They are per-pixel loss functions used in feed-forward image transformations.  In SRGAN \cite{Ledig2016} and its variants, three loss functions are used.  Empirically, none of the loss functions among the three can generate a convincing image alone.  In our study, we use at least two losses in various combinations 
to determine if and how they contribute to the overall outcome. 

\begin{itemize}

\item Image Loss (I): also referred to as MSE loss in Ledig et al. \cite{Ledig2016}.  This is a pixel-wise L2 loss between the output of the generator and the ground truth.  We call it image loss in order to distinguish it from the mean squared error that is used in VGG loss.  The resources required to calculate the image loss are the least expensive among the three.  We used L2 in this paper, however it is also possible to use L1.

\begin{equation}
L(G) = \sum_{x = 1}^W\sum_{y = 1}^H(I_{x, y}^{target} - G(I^{Input})_{x, y})^2
\end{equation}

\item VGG loss (V): Using a pre-trained VGG Network \cite{vgg2015} (available in \cite{urlPytorchModels}) plays a crucial role for training the generator.  The network has been trained with ImageNet and already knows what real-world images look like, delivering the results for object classification/identification as well as synthesis.  Given $\phi$ is a VGG network, the loss function is defined to be

\begin{equation}
\begin{split}
L(G) = \sum_{x = 1}^W\sum_{y = 1}^H[ \phi_{i, j}(I^{target})_{x, y} \\ - \phi_{i, j}(G(I^{Input}))_{x, y} ]^2
\end{split}
\end{equation}

\noindent
where $\phi_{i,j}$ refers to the feature maps obtained from the j-th Convolution/ReLU pair before the i-th maxpooling layer within the VGG-19 network, the same notation used in \cite{Ledig2016, Wang2018}.

\item Adversarial Loss (A): which is calculated with the discriminator, is what makes the system a GAN.  In other words, in the absence of this loss, there is no need for a discriminator, and the resulting framework is no longer designated as a generative adversarial network.  The resources required for computing the adversarial loss and how impactful it is in our image reconstruction deserves attention.

\begin{equation}
L(G) = \sum_{n = 1}^N - \log D(G(I^{Input}))
\end{equation}

\end{itemize}

\vspace{12pt}
\noindent
The total loss, $L$, is calculated as

\begin{equation}
\label{eq1}
L = \lambda_0I + \lambda_1V + \lambda_2A
\end{equation}
where $I, V, A$ represent image loss, VGG loss, and adversarial loss, respectively.  In the actual calculation, we set $\lambda_0 = 1$, so that only two parameters $\lambda_1$ and $\lambda_2$ are considered to determine the portion of each loss influencing the computation.  In Sections \ref{secAdvLoss} and \ref{secVGG}, we examine these losses more closely.

%%%%%%%%%%%%%%%%%%%%%%%%%%%%%%%%%%%%%%%%%%%%%%%%%%%%%%%%%%%%%%%%%%%%%%%%%%%%%%%%
\section{Experiment 1.\\-- Impact of Adversarial Loss}
%%%%%%%%%%%%%%%%%%%%%%%%%%%%%%%%%%%%%%%%%%%%%%%%%%%%%%%%%%%%%%%%%%%%%%%%%%%%%%%%
\label{secAdvLoss}

In the realm of supervised training, there are quite a few papers that report successful reconstruction of images without adversarial loss \cite{ChenKoltun, Kim2015AccurateIS, Lim2017, Tong2017, Mei2019DeepRR, Ma2019AMA}, which indicate that adversarial loss is not an essential component for image reconstruction.  Prior to the architecture of GFE described in Section \ref{secGFE}, this section analyzes the value of adversarial loss with the degree of its effect on both super-resolution (SR) and canny edge (Canny).  In this experiment, we used smaller image sizes as well as a shallower network than those used in the experiments 2 and 3.

%This is true not only for single image super-resolution, but also for GFE. %

\subsection{Architecture}
\label{secAdvLossArch}

The generator consists of 16 layers of residual blocks, each with 64 feature maps.  This is the structure used in \cite{Ledig2016}.  We used it for both SISR and image reconstruction from edges (Canny) for the experiment.  The discriminator has eight convolutional layers with an increasing number of feature maps; 64-64-128-128-256-256-512-512, followed by two dense layers and a sigmoid activation function.  In search of a suitable implementation we turn off VGG loss, if any, and run only in adversarial loss to see how the network converges.  We selected a few implementations published in Github \cite{urlSrez, urlTwhui} among those that converge with adversarial loss only, and plugged them into our implementation so that fair comparisons can be made.  The sizes of input and output images are the same; we experimented on 3 sizes -- $96 \times 96$, $128 \times 128$, and $176 \times 176$ for both SR and Canny.

\subsection{Methods}
Since the image loss has the minimum overhead to calculate, we leave it in all three scenarios listed below.  In all three cases, we set $\lambda_0 = 1$ in Equation (\ref{eq1}).  We ran 20 epochs and took the best Fréchet Inception Distance (FID) \cite{Martin2017, urlTTUR}.  FID uses a pre-trained Inception network and calculates the Fréchet distance between two multivariate Gaussian distributions with mean $\mu$ and covariance $\Sigma$,

\vspace{3pt}
\noindent
FID$(x, g) = ||\mu_x - \mu_g||^2 +$ Tr$(\Sigma_x + \Sigma_g - 2(\Sigma_x \Sigma_g)^{1/2})$

\vspace{3pt}
\noindent
where $x$, $g$ are the activations of the pool\_3 layer of the Inception-v3 net for real samples and generated samples, respectively.\\

%\begin{itemize}
%  \setlength{\itemsep}{0pt}
\begin{compactitem}
\item Image loss + VGG loss ( I + V)\\
  \hspace{10pt} ($\lambda_1 > 0$ and $\lambda_2 = 0$)
\vspace{6pt}
\item Image loss + Adversarial loss (I + A)\\
  \hspace{10pt} ($\lambda_1 = 0$ and $\lambda_2 > 0$)
\vspace{6pt}
\item Image loss + VGG loss + Adversarial loss \\
  \hspace{10pt}(I + V + A) ($\lambda_1 > 0$ and $\lambda_2 > 0$)
%\end{itemize}
\end{compactitem}

%\begin{figure}[H]
%\centering
%\subfloat[] { \includegraphics[width=0.48\linewidth]{Figures/IVA_SR.png} \label{IVA_SR}}
%\subfloat[] { \includegraphics[width=0.48\linewidth]{Figures/IVA_Canny.png} \label{IVA_Canny}}
%\caption{FID Scores for combinations of losses (lower is better) (a) SR (b) Canny}
%\label{figIVA}
%\end{figure}

\begin{figure}[H]
\centering
\includegraphics[width=0.9\linewidth]{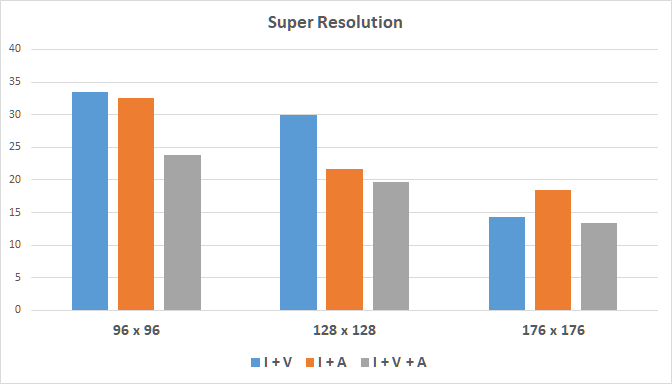}
\caption{FID Scores -- combinations of losses for SR (lower is better) \label{IVA_SR}}
\label{figIVA}
\end{figure}

\begin{figure}[H]
\centering
\includegraphics[width=0.9\linewidth]{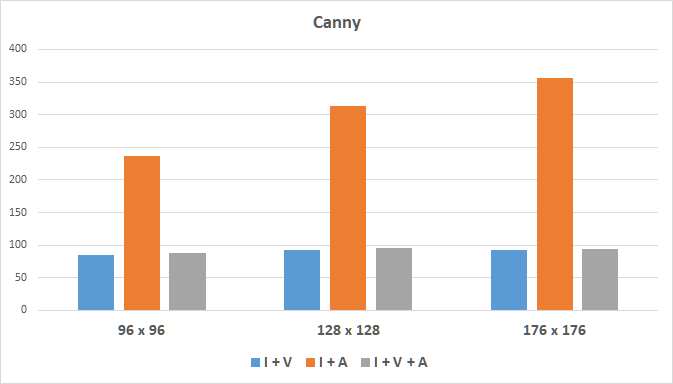}
\caption{FID Scores -- combinations of losses for Canny (lower is better) \label{IVA_Canny}}
\label{figIVA}
\end{figure}

%(a) I+V: $\lambda_1 = 0.001$, I+A: $\lambda_2 = 0.01$, I+A+V: $\lambda_1 = 0.001, \lambda_2 = 0.005$\\
%(b) I+V: $\lambda_1 = 0.001$, I+A: $\lambda_2 = 0.01$, I+A+V: $\lambda_1 = 0.001, \lambda_2 = 0.01$\\
%(c) I+V: $\lambda_1 = 0.005$, I+A: $\lambda_2 = 0.005$, I+A+V: $\lambda_1 = 0.001, \lambda_2 = 0.005$\\
%(d) I+V: $\lambda_1 = 0.01$, I+A: $\lambda_2 = 0.01$, I+A+V: $\lambda_1 = 0.005, \lambda_2 = 0.01$ }

\subsection{Discussion}
\label{advLossDiscussion}
As commonly seen, the more complex the task is, the more difficult it is for the generative adversarial network to converge.  For SISR, Figure \ref{IVA_SR} clearly shows the contribution by adversarial loss to the image quality, especially in lower resolutions.  However, for synthesis from canny images, a task more complex than SR,  adversarial loss does not show any positive contribution to the outcome.  In fact, we could not successfully generate convincing images at all if VGG loss is not included (the case [I + A] in Figure \ref{IVA_Canny}).  SISR is easier for image reconstruction, where adversarial loss can be incorporated into part of the perceptual loss more naturally than the reconstruction from canny edges.

\begin{figure}[H]
\centering
\subfloat[] { \includegraphics[width=0.48\linewidth]{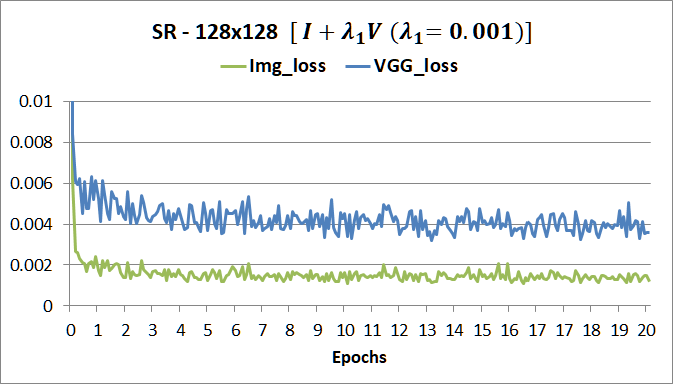} \label{SR128_V}}
\subfloat[] { \includegraphics[width=0.48\linewidth]{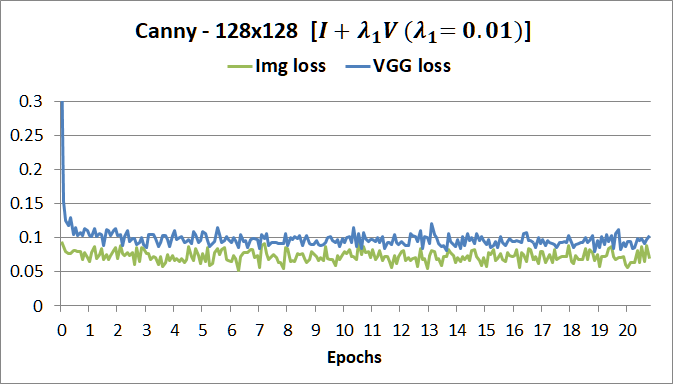} \label{Canny128_V}}\\
\subfloat[] { \includegraphics[width=0.48\linewidth]{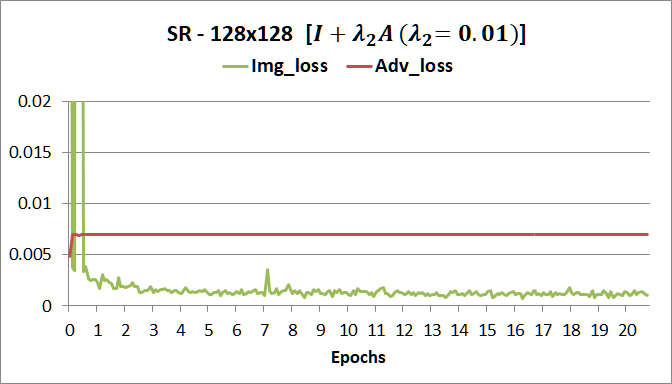} \label{SR128_A}}
\subfloat[] { \includegraphics[width=0.48\linewidth]{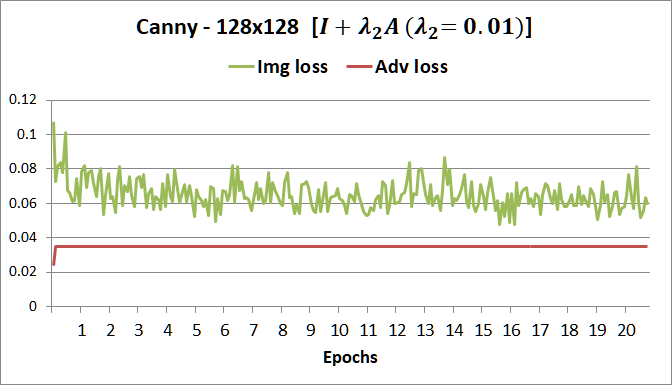} \label{Canny128_A}}\\
\subfloat[] { \includegraphics[width=0.48\linewidth]{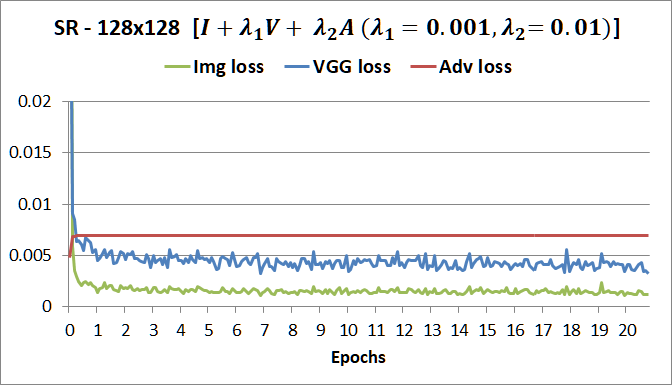} \label{SR128_VA}}
\subfloat[] { \includegraphics[width=0.48\linewidth]{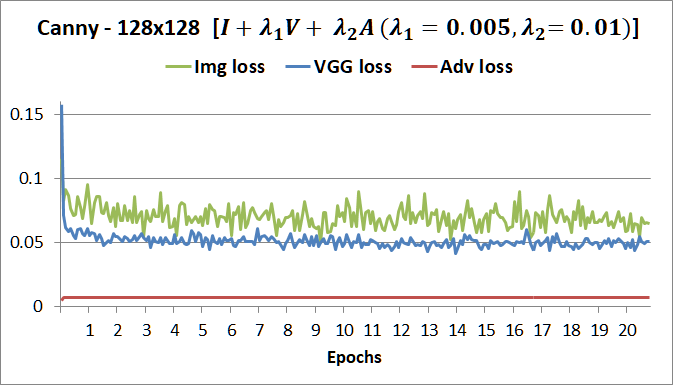} \label{Canny128_VA}}
\caption{Loss values by epochs from Figure \ref{IVA_SR} [$128 \times 128$]\\ (a) I+V (c) I+A (e) I+V+A,\\ and Loss values from Figure \ref{IVA_Canny} [$128 \times 128$]\\(b) I+V (d) I+A (f) I+V+A}
\label{SRCanny128_3}
\end{figure}

We recorded the loss values as the training continued at each epoch.  Figure \ref{SRCanny128_3} shows sample loss values over the course of training for the size of $128 \times 128$ of Figures \ref{IVA_SR} and \ref{IVA_Canny}.  While image loss and VGG loss show a typical, oscillating yet steady decrease in values, adversarial loss converges rather quickly to a constant value after several hundred iterations.  This raises a few interesting theoretical points: First, if we knew the constant value in advance, we could use it in lieu of the adversarial loss and save computer resources.  Second, if we could come up with a method to decrease the adversarial loss throughout the training, we could take full advantage of the power of the generative adversarial network.  For now, however, these are left for future research, and we conclude that adversarial loss does not contribute to the synthesis of images from canny edge, and that the resource is better used for a larger generator.  Consequently at this point, GAN is not used in our study.  Unless otherwise noted, the remainder of this paper uses only image loss and VGG loss.

%%%%%%%%%%%%%%%%%%%%%%%%%%%%%%%%%%%%%%%%%%%%%%%%%%%%%%%%%%%%%%%%%%%%%%%%%%%%%%%%
\section{Experiment 2.\\-- Optimizing VGG Loss}
%%%%%%%%%%%%%%%%%%%%%%%%%%%%%%%%%%%%%%%%%%%%%%%%%%%%%%%%%%%%%%%%%%%%%%%%%%%%%%%%
\label{secVGG}

VGG-19 consists of 16 Convolution layers, each followed by ReLU activation.  Between the last (16th) ReLU layer and the output softmax layer, there are three fully connected layers, which are not used for perceptual loss.  For perceptual loss, both aforementioned papers used the last pair (16th layer, $\phi_{5,4}$); while Ledig et al. \cite{Ledig2016} used the activation layer, Wang et al. \cite{Wang2018} claims it is more effective to use the convolutional layer before the activation, which we confirm to be true.  In this experiment, we further analyze using VGG loss computed from various layers and recommend an optimized VGG loss for  our image reconstruction application.

\subsection{Architecture and Dataset}
\label{secVGG_sub1}
We used VGG-19 along with image loss as part of the perceptual loss in the generator (GFE).  As we established in Section \ref{advLossDiscussion}, adversarial loss is not used.  Consequently we can eliminate the discriminator.  The architecture of the generator is the same as the one used in Section \ref{secAdvLoss}, but with the CelebA-HQ dataset --- it consists of 30,000 high-resolution images with the size $1024 \times 1024$.  We resize them to $224 \times 224$, which is a size required by Section \ref{resnet-loss}.  Removing 2 outliers (imgHQ00070 and imgHQ02815), and setting aside 1,000 images for validation/testing, we have 28,998 images for training.

\subsection{Result}
\begin{figure}[H]
\centering
\includegraphics[width=1\linewidth]{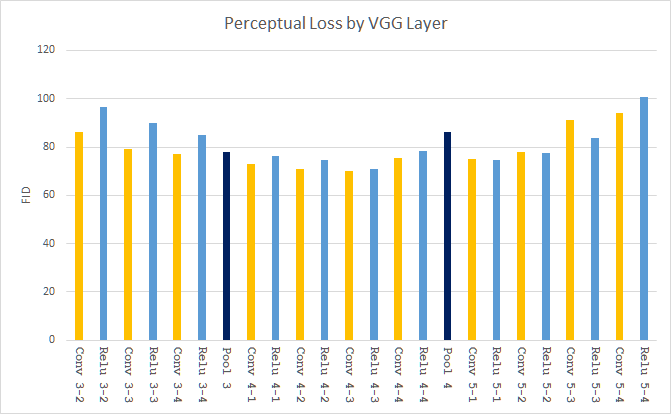}
\caption{VGG-19 -- layer by layer analysis (lower is better) --- Yellow bars are convolutional, Blue is ReLU, and Black is Pool Layer}
\label{vgg19}
\end{figure}

Contrary to common usage of how VGG loss is applied, our study shows using middle layers is more effective than using later layers, either the convolutional layer or the activation layer.  Discarding the later layers also saves the memory space in the hardware.  
Figure \ref{vgg19} shows that the convolutional layers of $\phi_{4,2}$ and $\phi_{4,3}$, (10th and 11th convolutional layers, respectively) show the best FID scores.

\subsection{Multiple layers of VGG loss}

\begin{figure}[H]
\centering
\includegraphics[width=.9\linewidth]{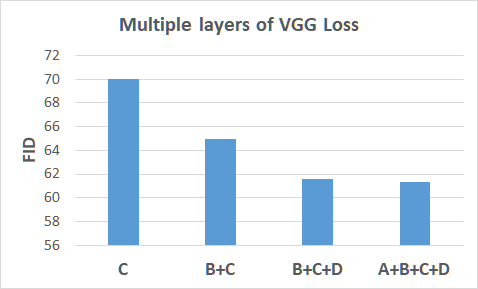}
\caption{[A] Conv4-1;  [B] Conv4-2;  [C] Conv4-3;  [D] ReLu4-3, with 16 block layers of generator (lower is better)}
\label{figMultiVGG}
\end{figure}

More than one layer can be used as part of perceptual loss.  Without the assistance of adversarial loss, we have $\lambda_2 = 0$, and assuming $\lambda_0 = 1$, Equation (\ref{eq1}) becomes

\begin{equation}
L = I + \lambda_{1_1}V + \lambda_{1_2}V + \dots + \lambda_{1_n}V
\end{equation}

\noindent
where $n$ is the number of VGG layers to be used for perceptual loss.  Starting with Conv $\phi_{4,3}$, we selected the best 4 layers and added new layers one by one (see Fig. \ref{vgg19} for reference to layers).  Figure \ref{figMultiVGG} shows a sample of FID scores for $n = \{1, 2, 3, 4\}$ in our experiments.  Using two layers ($n = 2$) is better than using a single layer ($n = 1$), and $n = 3$ is better than $n = 2$.  But for $n > 3$, the effect of adding extra layers diminishes.  Most of our experiments in Section \ref{secGFE} use 2 layers of Conv $\phi_{4,2}$ and Conv $\phi_{4,3}$.

\subsection{Resnet Loss}
\label{resnet-loss}
\begin{figure}[H]
\centering
\includegraphics[width=.9\linewidth]{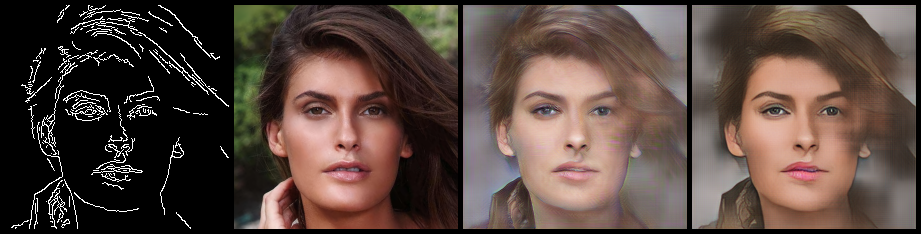}\\
\includegraphics[width=.9\linewidth]{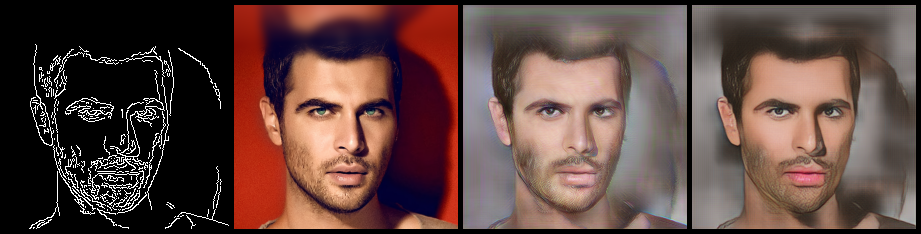}
\caption{From Left: Input, GT, VGG, Resnet}
\label{figResnet}
\end{figure}

Instead of VGG loss, it is possible to use other models as a part of perceptual loss.  In this subsection, we experimented with a pre-trained Resnet152 model \cite{urlPytorchModels}.  Unlike VGG, however, the Resnet model consists of blocks of ``Bottleneck", and we were able to detach only the outer-most layers.  Nevertheless, the resulting images exhibit unique characteristics that are not present in using VGG alone.  Samples are shown in Figure \ref{figResnet} where two Resnet layers are used.  Images in column 4 show that fine details are omitted and the skin of each face appears too smooth.  We concluded we cannot improve the image quality further using the same method by larger Resnet networks, and determined that unless we go deep inside Bottleneck blocks, it is as far as we can go.

%\end{multicols}

%%\newpage
%\begin{multicols}{1}
%\begin{figure}[H]
%\centering
%\hspace*{0.4cm}
%\subfloat {\includegraphics[width=.45\textwidth]{Sketches/fourImages_Sketches_f1-001-01-sz1_150-300.png}}
%\hspace*{0.3cm}
%\subfloat {\includegraphics[width=.45\textwidth]{Sketches/fourImages_Sketches_m1-001-01-sz1_150-300.png}}\\
%%\vspace{1pt}
%\subfloat {\includegraphics[width=.33\textwidth]{Sketches/threeImages_Sketches_1776cc2e00ac888aadcc2fce71552020_30-70.png}}
%\hspace*{.1cm}
%\subfloat {\includegraphics[width=.33\textwidth]{Sketches/threeImages_Sketches_f1a3e780e2e589b06e15891c09c8d480_50-150.png}}
%\hspace*{.1cm}
%\subfloat {\includegraphics[width=.33\textwidth]{Sketches/threeImages_Sketches_face2_150-300.png}}
%%\caption{From Left: (Top) GT, Sketch, Canny, Output (Bottom) Sketch, Canny, Output}
%\captionof{figure}{From Left: (Top) GT, Sketch, Canny, Output \\(Bottom) Sketch, Canny, Output}
%\label{figSketches}
%\end{figure}
%\end{multicols}

%\begin{multicols}{2}

%%%%%%%%%%%%%%%%%%%%%%%%%%%%%%%%%%%%%%%%%%%%%%%%%%%%%%%%%%%%%%%%%%%%%%%%%%%%%%%%
\section{Experiment 3.\\-- Generator From Edges}
%%%%%%%%%%%%%%%%%%%%%%%%%%%%%%%%%%%%%%%%%%%%%%%%%%%%%%%%%%%%%%%%%%%%%%%%%%%%%%%%
\label{secGFE}

We form GFE based on the results obtained from Sections \ref{secAdvLoss} and \ref{secVGG}.  Increasing the size of the network is effective up to a certain point due to the vanishing gradient problem, and residual blocks along with skip connections are notable solutions for large networks \cite{He2015DeepRL}.  Recently in SISR, studies claim that making the residual block denser (more connections within the block), as well as having more skip connections between blocks improves performance.  For GFE however, these claims are not applicable.  In this section, we describe the details of experiments in pursuit of the best architecture for GFE.

Given the same number of parameters, empirically a deep neural network (more layers) is more efficient than a wide network (more feature maps).  Many networks are constructed in such a way that as the layers go deeper at later stages, the number of feature maps increases until just before the fully connected layers.  On the other hand, except for the last block layers, SRGAN \cite{Ledig2016} has a constant number of feature maps (64) throughout the generator; a monolithic architecture.  We adopt this strategy.

\subsection{Architecture}

We explore ideas proposed in the field of SISR for our application, mainly dense blocks and skip connections.  By using a fixed number of feature maps at every block layer, we can focus on the study of structures in the residual block and skip connections for how dense and how deep the network should be.  The number of feature maps at each layer is 64, and the kernel sizes are all $3 \times 3$.  Starting with 16, we increase the number of block layers at increments of 8.  Without the discriminator, we have more memory available for the construction of GFE.  All experiments were conducted in a single GPU with 11GB of memory, and it is worth noting that the image size we generate ($224 \times 224$) is mainly determined by the capacity of the GPU memory for training.  The same dataset as Section \ref{secVGG} (Experiment 2) is used.

\subsection{Sketch to Photo}

\begin{figure}[H]
\centering
\hspace*{0.4cm}
\subfloat {\includegraphics[width=.45\textwidth]{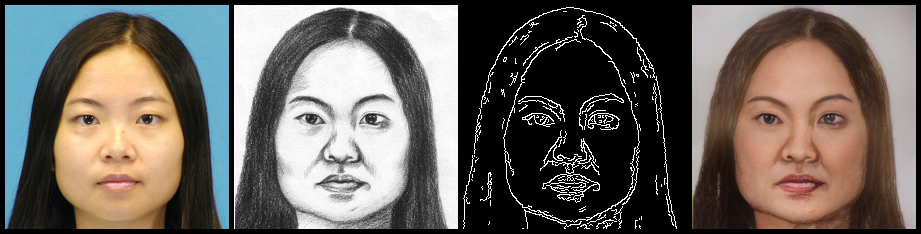}}
\hspace*{0.3cm}
\subfloat {\includegraphics[width=.45\textwidth]{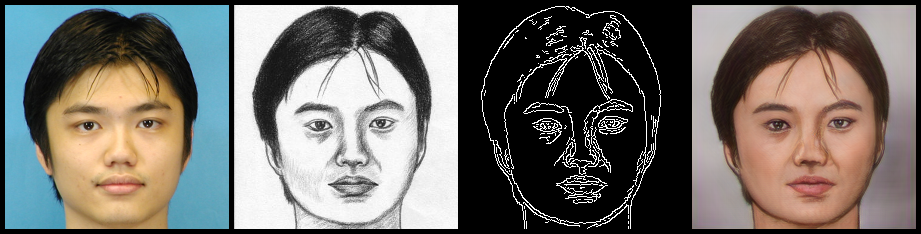}}\\
\vspace{1pt}
\subfloat {\includegraphics[width=.33\textwidth]{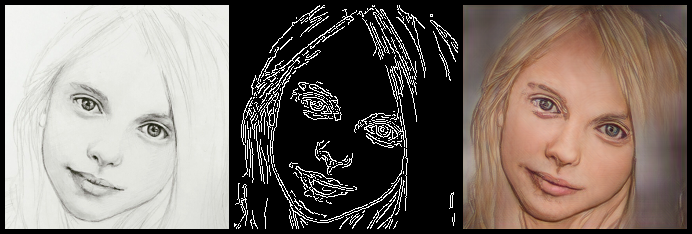}}
\hspace*{.1cm}
\subfloat {\includegraphics[width=.33\textwidth]{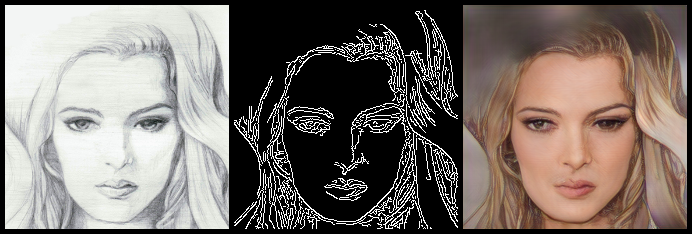}}
\hspace*{.1cm}
\subfloat {\includegraphics[width=.33\textwidth]{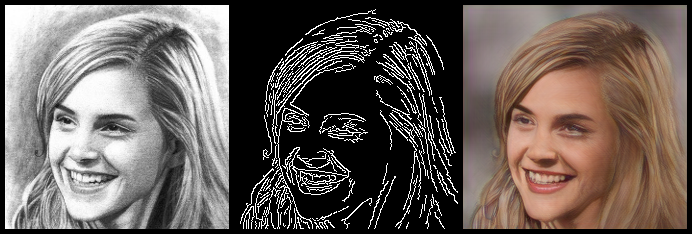}}
%%%\caption{From Left: (Top) GT, Sketch, Canny, Output (Bottom) Sketch, Canny, Output}
\captionof{figure}{From Left: (Top) GT, Sketch, Canny, Output \\(Bottom) Sketch, Canny, Output}
\label{figSketches}
\end{figure}

Figure \ref{figSketches} shows some potential practical applications with GFE. We took some pencil sketches from CUHK \cite{urlCUHK} as well as from the internet.  Note that since we trained with the canny transformation, we first have to convert the sketch image to canny, and then make an inference with the trained network.  Not just as a simple coloring exercise, in the output images we can see the depth of textures of human face that commonly appear in every person, which shows a great potential for an image translation from a pencil sketch to a photo.

\end{multicols}
%\newpage
\begin{multicols}{1}
\end{multicols}
\begin{multicols}{2}

\subsection{Residual Block}
\label{secResidualBlock}

\begin{figure}[H]
\centering
\includegraphics[width=.5\linewidth]{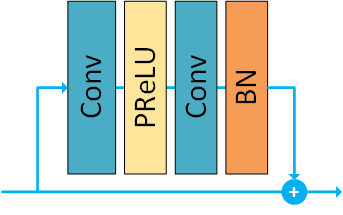}
\caption{Residual Block (base unit)}
\label{figBaseBlock}
\end{figure}

The base unit of the construction, often called the residual block, is illustrated in Figure \ref{figBaseBlock}.  The input is followed by a convolutional layer, followed by a ParametricReLU and another convolutional layer.  Then a batch normalization (BN) is added before the output that is combined with the input as a single dense connection.  This is very similar to SRGAN \cite{Ledig2016}; the difference being the omission of the first BN layer.  This omission is crucial for reducing the memory footprint.

The batch normalization layers consume the same amount of memory as the preceding convolutional layers, and removing a BN layer from the unit block saves us approximately 20\% of the memory space in our model.  If we had removed both BN layers, we would have saved 40\% of the memory usage \cite{Lim2017}, but our experiments show that leaving in one BN layer yields better results than none at all.  Comparing a single BN layer with two BN layers, we found no noticeable differences.

Several studies in SISR propose dense residual blocks \cite{Wang2018, Tong2017}, but a generator with such dense residual units requires considerably more GPU memory, forcing us to train the network with smaller batch sizes (mini-batches).  The dense blocks contribute to a better outcome for a network with a small number of block/unit layers.  However, when we form a larger network, such dense units negatively impact the result. For a generator with 32 or more block layers, dense residual blocks are inadequate.

Thus, we use one connection within the block, between the input and the BN layer.  Even with just one connection at each block, when a generator is constructed by having residual blocks stacked multiple times, the entire network is connected in such a way that the gradient vanishing problem is dramatically reduced, and the outcome is significantly better than without using residual blocks.

\subsection{Skip Connections and Large Networks}

\begin{figure*}[ht] %[ht] = top
\hspace{1.2cm}
\includegraphics[width=.8\linewidth]{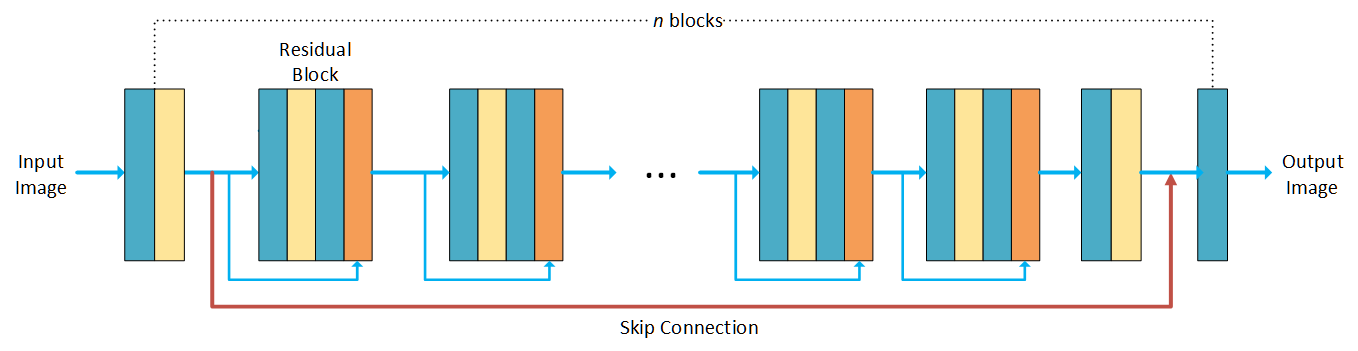}
\caption{Skip Connection Type 1}
\label{figSkipConnections}
\end{figure*}

Let the number of basic blocks (residual blocks in the middle, and blocks for Conv + PReLU at the beginning and end of the network) be $n$ [Figure \ref{figSkipConnections}].  We define the skip connection type as follows:

\begin{itemize}
  \setlength{\itemsep}{0pt}
%\begin{compactitem}
  \item Type 0: No skip connection
  \item Type 1: Connect with layer 1 and layer $n - 1$
  \item Type 2: Connect with layer 1 and layer $n / 2$, as well as $n / 2$ and $n -1$
  \item Type 3: Connect with layer 1 and layer $n - 1$, as well as layer $n / 2$ and $n -1$
  \item Type 4: Connect with Type 1 and Type 2 combined
\end{itemize}
%\end{compactitem}

\begin{figure}[H]
\includegraphics[width=1\linewidth]{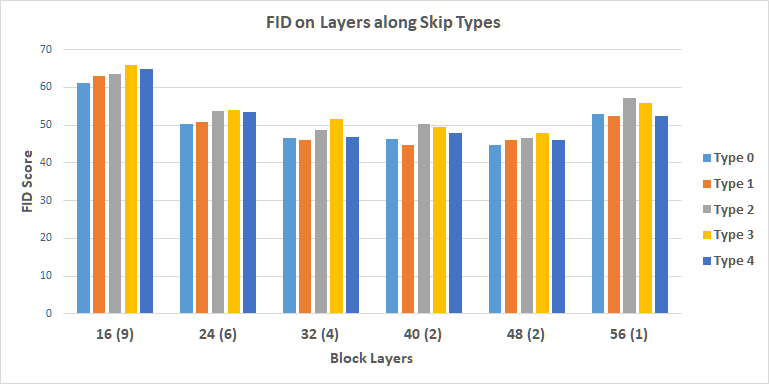}
\caption{Deep Networks with each type of skip connections (smaller is better; numbers in parentheses indicate batch size)}
\label{figDeepNetworks}
%\end{figure*}
\end{figure}

\noindent
Figure \ref{figSkipConnections} shows skip connection Type 1.  By going deeper in the generator, the output of synthesized images becomes better, and we found that forming 48 block layers (with each block consisting of 4 layers [Figure \ref{figBaseBlock}]) achieves the best result.  We tested with the above 5 skip connections to see which type is best using the residual block defined in \ref{secResidualBlock}.  Again, contrary to claims that having more skip connections improves image quality, none of the connection types has a positive effect for our application [Figure \ref{figDeepNetworks}].  While we conclude that no skip connection is necessary for our monolithic architecture of GFE, the combination of residual block density and how to use skip connections among the blocks in other forms of architecture is another area of study to be explored in the future, including those proposed by \cite{Lim2017, Mei2019DeepRR, Ma2019AMA}.

%\begin{figure*}[ht] %[ht] = top
%\includegraphics[width=1\linewidth]{Figures/SkipConnections-lrg.png}
%\caption{Skip Connection Type 1}
%\label{figSkipConnections}
%\end{figure*}

\subsection{The limit of Depth in the generator}

There is a limit on how deep we can construct a network for the best outcome.  Larger networks are not necessarily better than smaller ones.  We started out our experiment with 16 layers of residual blocks, with a batch size of 9 (nine images are processed in the GPU at once in a single iteration).  As we increased layers, we had to decrease the batch size due to the limitation of GPU memory. Initially the image quality improved but soon it saturated in improvement.  

We observed some degradation for a network whose block size is greater than 48, where the mini-batch size needs to be 1 (one) to fit in our GPU.  At this point we suspect that batch normalization is no longer in effect, and in fact the training is somewhat unstable (consecutive epochs have FID values in a wide swing).  Although we attempted to tweak hyper parameters such as learning rate, we were unable to improve image quality.

%\begin{figure*}[ht]

\subsection{Adam Optimizer and L2 Loss}
We used an Adam optimizer in all our experiments.  In the Pytorch implementation of Adam, the user can set an initial learning rate as a parameter, but the learning rate decay is not provided.  Although the authors claim that parameter updates are invariant \cite{Adam}, we found that a decay rate of 0.02 per epoch gives the best result, and as such we manually updated the learning rate with an initial rate of 0.0005.  All the remaining parameters were left at the default settings.

We used L2 (MSE) loss for perceptual loss calculation throughout our study.  Despite certain claims that L1 loss gives a better result, in our experiments FID scores are consistently better using L2.

%%%%%%%%%%%%%%%%%%%%%%%%%%%%%%%%%%%%%%%%%%%%%%%%%%%%%%%%%%%%%%%%%%%%%%%%%%%%%%%%
\section{Conclusion and Discussion}
%%%%%%%%%%%%%%%%%%%%%%%%%%%%%%%%%%%%%%%%%%%%%%%%%%%%%%%%%%%%%%%%%%%%%%%%%%%%%%%%

We demonstrated the Generator From Edges (GFE) for image translation on human faces, from edges to photo, without a generative adversarial network. This was led by the analysis of architectural features that unnecessarily consume GPU memory, such as a discriminator and extra batch normalization layers.  We also reviewed a dense network and observed that skip connections are not effective if the basic unit is densely connected.  As the generator becomes larger, the image quality improves, but there is a point beyond which a larger network size is ineffective.

Although the trained network can restore facial images even when the edges are not drawn precisely in the input \cite{NaoGita}, the nature of supervised training commands deterministic outputs.  For a practical application in mind, however, removing the GAN loses stochasticity in the inference mechanism, in which when an incomplete image is fed to the network, the outcome would also be less than ideal.  This could be addressed with an unsupervised training in such a way that incomplete input leads to more convincing output.  At the same time, as mentioned in Section \ref{secRelatedWorks}, we are seeing rapid advancements in research --- such as mask-guided (with GAN) or geometry-guided (with VAE) settings --- to fill in the gap where nondeterministic outcomes are desired.

\bibliography{References}
\bibliographystyle{unsrt}

\end{multicols}
\end{document}